\documentclass[aps,prx,twocolumn,groupedaddress,showpacs,amsmath,amssymb,longbibliography, amsfonts,10pt]{revtex4-2}

\usepackage{graphicx} 
\usepackage{amsmath}
\usepackage{amsfonts}
\usepackage{amssymb}
\usepackage{graphicx}
\usepackage{color}
\usepackage{xcolor}
\usepackage{bbold}
\usepackage{appendix}
\usepackage[citecolor=blue]{hyperref}
\hypersetup{colorlinks}
\usepackage[caption=false]{subfig}
\usepackage{bookmark}
\usepackage{makecell}
\usepackage{mathtools}
\usepackage{comment}
\usepackage{tikz}
\usepackage{pgfplots}
\usetikzlibrary{arrows.meta}
\usepackage{array}
\usepackage{multirow}
\usepackage[nolist,nohyperlinks]{acronym}
\usepackage[export]{adjustbox}
\usepackage{placeins}
\usepackage{soul}
\pgfplotsset{compat=1.18}

\newacro{CEP}[CEP]{critical exceptional point}
\newacro{frg}[FRG]{functional RG}
\newacro{DSE}[DSE]{Dyson-Schwinger equation}
\newacro{MSRJD}[MSRJD]{Martin-Siggia-Rose-Janssen-De Dominicis}
\newacro{BKT}[BKT]{Berezinskii–Kosterlitz–Thouless}
\newacro{KPZ}[KPZ]{Kardar–Parisi–Zhang}
\newacro{EFT}[EFT]{effective field theory}
\newacro{Nlsm}[NL$\sigma$M]{non-linear $\sigma$ model}

\newcommand{\vecq}{\boldsymbol{q}}
\newcommand{\xx}{\boldsymbol{x}}

\newcommand{\vv}[1]{\boldsymbol{#1}}

\newcommand{\vphi}{\boldsymbol{\phi}}
\newcommand{\vvarphi}{\boldsymbol{\varphi}}

\newcommand{\veca}{\boldsymbol{a}}

\newcommand{\expVal}[1]{\left\langle #1 \right\rangle}

\graphicspath{{figures/}}
\begin{document}

\title{Kardar-Parisi-Zhang scaling in time-crystalline matter}
\author{Romain Daviet}
\thanks{These authors contributed equally to this work.\\  Emails: daviet@thp.uni-koeln.de   zelle@thp.uni-koeln.de}
\author{Carl Philipp Zelle}
\thanks{These authors contributed equally to this work.\\  Emails: daviet@thp.uni-koeln.de   zelle@thp.uni-koeln.de}
\author{Armin Asadollahi}

\author{Sebastian Diehl}

\affiliation{Institut f\"ur Theoretische Physik, Universit\"at zu K\"oln, 50937 Cologne, Germany}
\begin{abstract}
    We discuss the universal behavior linked to the Goldstone mode associated with the spontaneous breaking of time-translation symmetry in many-body systems, in which the order parameter traces out a limit cycle. We show that this universal behavior is closely tied to Kardar-Parisi-Zhang physics, which can strongly affect the scaling properties in all dimensions. Our work predicts an rationalizes the emergence of KPZ in numerous systems such as nonreciprocal phases in active matter, active magnets, driven-dissipative quantum systems, and synchronization of oscillators. 
\end{abstract}
\maketitle

\paragraph*{Introduction -- } Time-translation symmetry is a fundamental symmetry of matter in and out of equilibrium. Breaking it spontaneously is impossible in thermodynamic equilibrium governed by a bounded Hamiltonian \cite{Bruno2013,Nozieres2013,Volovik2013,Watanabe2015}, but it occur both in quantum and classical systems far from equilibrium. A particular incarnation are dynamical limit cycles -- or continuous time crystals -- where the macroscopic order parameter retains a periodic motion. This phenomenon has recently gained much attention experimentally as well as theoretically in engineered quantum systems, ranging from ultracold Bose condensates immersed into optical cavities \cite{Piazza2015,owen2018,Dogra2019,Buca2022,Buca2019} over dissipative Bose condensates such as exciton-polariton systems \cite{Carusotto2013,Sieberer2013,Hanai2019}, magnon condensates~\cite{Demokritov2006,Rezende2009,NowikBoltyk2012, Bunkov2013,Autti2018,Autti2020}, and continuous dissipative time crystals~\cite{Iemini2018,Buca2019,Scarlatella2019,Kessler2019,Kongkhambut2022}. But limit cycles in many-body systems are also relevant in classical systems~\cite{Cross1993,Brunnet1994,Risler2005,Guislain2023,Guislain2024,Meibohm2024}, in particular through synchronization of oscillators relevant in many different contexts, including in biology and chemistry~\cite{Kuramoto1984,Pikovsky2001}. Recently,  realizations of such phases have gained a lot of interest in active matter set-ups, for example in systems with nonreciprocal interactions \cite{Fruchart2021,Avni2023,Zelle2024,Saha2020,You2020,FrohoffHuelsmann2021,Brauns2024,Suchanek2023a,Hanai2024,Guislain2024a,Cates2024}, chiral rotating states~\cite{Maitra2020,Maitra2024} and also through generalization to `active' magnetic systems \cite{delSer2021,delSer2023,Hanai2024a,hardt2024}.

Since time translations form a continuous symmetry, breaking it spontaneously gives rise to a gapless mode as a consequence of the Goldstone theorem \cite{Hayata2018,hongo2019,Zelle2024,Daviet2024,denova2024}. Because time-translation invariance is generically an exact symmetry, this Goldstone mode must be a robust origin of universal scaling throughout extended regions of parameter space in a plethora of nonequilibrium systems. In this work, we distill the universal behavior connected to spontaneous time-translation symmetry breaking in driven open systems, with an emphasis on limit-cycle phases. 
Two key factors determine their universal physics. First, the limit cycle periodicity itself dictates that the Goldstone mode of time translations~\cite{Chan2015,hongo2019,Zelle2024} is an angular variable. Second, the necessarily nonequilibrium character of the underlying dynamics, together with the continuous growth of the limit-cycle phase, implies that this Goldstone mode is subject to the non-linearity of the paradigmatic Kardar-Parisi-Zhang (KPZ) equation \cite{Kardar1986,Krug1997,Takeuchi2018}. The Goldstone mode dynamics is thus governed by a compact KPZ equation~\cite{Chen2013,Altman2015,He2015,Lauter2017,Maitra2020}. The universal scaling behavior is then expected to show both a regime governed by the KPZ universality class, but also regimes dominated by the presence of vortex defects allowed by the compactness of the limit-cycle variable~\cite{Sieberer2016a,Wachtel2016,Vercesi2023,Deligiannis2022}. These theoretical findings are complemented by numerical simulations of extended Van der Pol systems in one and two spatial dimensions.

The emergence of the compact KPZ physics has been observed in numerous contexts, for example in classical time-dependent periodic states~\cite{Bennett1990,Grinstein1993,Chate1995,Manneville1996,Brunnet1998,Gutierrez2024,Gutierrez2024a}, but also as a realization through the phase variable of a broken quantum mechanical phase rotation symmetry~\cite{Altman2015,Sieberer2016}. This mechanism led to an experimental observation of the KPZ physics in one-dimensional exciton-polaritons~\cite{Fontaine2023}. Our work does not only reveal the common root behind these observations. The robust symmetry based mechanism also paves the way to systematically identify platforms that fall into the compact KPZ universality class.

\paragraph*{Time-translation symmetry breaking -- } A system is said to be time-translation invariant if its time evolution does not explicitly depend on time, i.e. it is not subject to time dependent forces or interactions. Time-translation symmetry is then broken spontaneously, if in the stable state there is an observable collective degree of freedom, that has a time-dependent expectation value $\expVal{\phi(t,\xx)}=\varphi(t)$, or equivalently 
\begin{align}
    \expVal{\phi(t,\xx)\phi(0,0)}\to \varphi(t)\varphi(0)\, \text{for} \, |\xx|,|t| \to \infty.
\end{align}
A time crystal thus necessitates a long-range order - as in usual symmetry-breaking scenarios -  of a coherent, time dependent periodic motion. Symmetry breaking is only possible in a thermodynamically large system, and typically occurs in dimensions larger or equal to three for continuous symmetries. However, in low dimensions, symmetry breaking and Goldstone modes are well known to be, in and out of equilibrium, fruitful starting points around which to expand in order to understand the long-wavelength physics~\cite{Altland2023,ZinnJustin,Beekman_2019}. 

Our approach encompasses both quantum and classical systems driven out of equilibrium. In the more general quantum case, by an open system Schwinger-Keldysh generating functional~\cite{Kamenev2011,Sieberer2016}, which takes the form (see supplemental material (SM) for details~\footnote{See Supplemental Material at [URL], which includes Refs.~\cite{Schwartz2014,Dupuis2022,Verhulst1996,Sun1989,Wolf1990}, for additional details on the mapping to the KPZ equation.})
\begin{align}\label{eq:keldysh}
    Z[\vv{\tilde{J}},\vv{J}]= \int \mathcal{D}\vphi \mathcal{D}\vv{\tilde\phi} \exp(iS[\vphi,\vv{\tilde\phi}]+i\vv{\tilde{J}} \vphi+i \vv{J} \vv{\tilde\phi}),
\end{align}
where $\vv{J},\vv{\tilde{J}}$ are source terms, which are needed to extract response and correlations functions, and $\vphi,\vv{\tilde\phi}$ are multicomponent fields, which can be chosen real without loss of generality~\footnote{In particular, for a complex order parameter, $\vphi$ collects its real and imaginary parts.}. $\vphi$ is the order parameter field, while $\vv{\tilde\phi}$ is its corresponding response field.
For the description of collective effects in generic driven open quantum systems, a semiclassical limit can be taken~\cite{Sieberer2016,sieberer2023universality}, in which the Schwinger-Keldysh action reduces to the Martin-Siggia-Rose-Janssen-De Dominicis action ~\cite{Martin1973,Janssen1976,DOMINICIS1976}. While still able to capture effects of quantum mechanical origin like phase coherence, this description is formally equivalent to classical stochastic Langevin dynamics.
 
\begin{figure}
    \centering
    \includegraphics[width=0.8\linewidth]{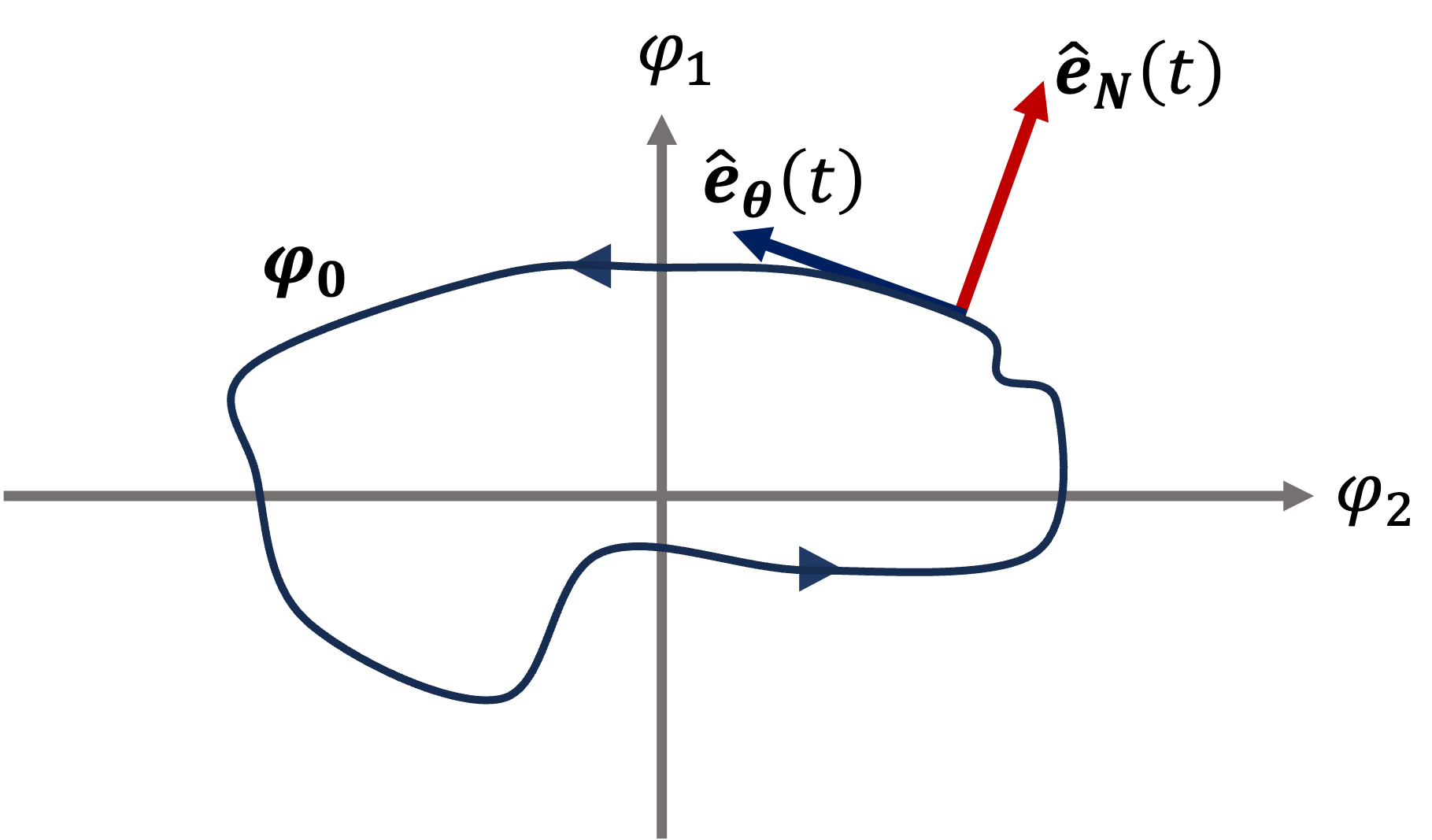} 
   
    \caption{Visualization of a generic limit cycle $\boldsymbol{\varphi}_0(t)$. The Goldstone mode $\theta(t,\xx)$ corresponds to fluctuations along the limit cycle in the direction $\hat{\boldsymbol{e}}_\theta=\dot\vvarphi_0(t)/ \lVert\dot\vvarphi_0(t) \rVert$, while $\hat{\boldsymbol{e}}_{N}(t,\xx)$ collects all remaining, directions perpendicular to the limit cycle, fluctuations along which are gapped. Note that in higher dimensional phase spaces, the limit cycle is still a one-dimensional closed curve. One can think of the chronon $\theta$ as the angle-like variable parametrising this curve. This does not exclude oscillations of single real degrees of freedom, as their phase space is constituted of $\vphi=(\phi,\partial_t\phi)^T$.}
    \label{fig:phase_space}
\end{figure}

In this formulation, a system is time-translation symmetric, if the microscopic action $S[\vphi,\vv{\tilde\phi}]$ does not depend on time explicitly but only implicitly through $\vphi(t),\vv{\tilde\phi}(t)$. 
The system spontaneously breaks (weak) time-translation symmetry, if the noise-averaged expectation value of the classical field is explicitly time dependent, $\partial_t\langle\vphi(t)\rangle=\partial_t\mathbf{\vvarphi}(t)\neq 0$ \footnote{In the Keldysh framework,  symmetries come with a fine structure often referred to as weak and strong or classical and quantum symmetries~\cite{Sieberer2016,sieberer2023universality}. In an open system, the strong time-translation symmetry is always broken (energy is not conserved), and the symmetry of interest is the weak version.}. In the case of a time crystal, $\expVal{\vphi(t)}$ traces out a closed orbit with period $T$, which can be parametrized by an angular variable $\theta_0\in[0,2\pi )$, $\vvarphi(t+\theta_0 \bar T)$ ($\bar T = \frac{T}{2\pi}$), see also ~Fig.~\ref{fig:phase_space}. Discrete time translations by $T$ remain unbroken. This is analogous to an ordinary crystal breaking continuous spatial translations down to discrete ones, but the time crystal is only possible out of equilibrium~\cite{Bruno2013,Watanabe2015}. 

In particular, if $\vvarphi(t)$ is a valid stable state — a solution of the equation of motion —  so is $\vvarphi (t+\theta_0 \bar T)$. Formally, we thus have a continuous manifold of equivalent stable states, with translations between the different states mediated by $\theta_0$. These transitions are gapless, since the Keldysh action for different $\theta_0$ is degenerate, constituting the soft Goldstone mode (see SM for a more formal derivation).  Again, the argument is analogous to the phonon in an ordinary crystal. We will refer to the Goldstone mode of broken time translations as the ``chronon''.

To construct the chronon excitations, according to the usual procedure for Goldstone modes, we now promote the global symmetry transformation to a local, slowly varying one: The constant time shift becomes $\theta_0 \rightarrow \theta(t,\xx) $, which varies on time scales much larger than the limit cycle period $T$. In total we can parametrize spatial fluctuations around the limit cycle within the path integral as
\begin{align}
    \vphi(t,\xx)
    =\vvarphi(t+\bar{T}\theta(t,\xx),\xx)+ \boldsymbol{N}(t,\xx), 
\end{align}
where $\theta(t,\xx)$ corresponds to local, gapless fluctuations \emph{along} the limit cycle and $\boldsymbol{N}(t,\xx)$ collects the gapped excitations perpendicular to it. Clearly, the chronon is represented by a compact $SO(2)$ Goldstone mode due to the limit cycle periodicity, see again Fig.~\ref{fig:phase_space} for a visualization.

\paragraph*{Effective field theory for the Goldstone mode --} To capture the effective dynamics of the chronon, we need to integrate out the gapped fluctuations $\boldsymbol{N}(t,\xx)$ and their respective noises. Before doing this explicitly, we construct the remaining action $S[\tilde\theta(t,\xx),\varphi(t+\theta(t,\xx)\bar{T})]$ based on symmetry. Since we are expanding around a time-dependent stable state, this will lead to an explicitly time-dependent action. Because $\theta(t,\xx)$ is fluctuating on timescales much larger than the period of the limit cycle $\tau\gg T$, we can eliminate these time-dependencies by averaging over a period $ T$. Afterwards, time-translation invariance of the original model implies invariance under a constant shift $\theta(t,\xx)+c$ and the effective field theory for the chronon is gapless, see SM for details~\cite{Note1}. To leading order, it has to take the form
\begin{align}
\label{eq:S_kpz}
    S=&\int_{t,\xx}\tilde\theta\left(\partial_t\theta-Z\nabla^2\theta+\frac{g}{2}(\nabla\theta)^2\right)-D\tilde\theta^2,
\end{align}
which is equivalent to the KPZ equation, as seen upon  passing from the MSRJD/Keldysh path integral to the  stochastic Langevin formulation \cite{Kamenev2023,Sieberer2016,sieberer2023universality}. 

The quadratic sector here describes a noisy, diffusive Goldstone mode.  The only symmetry allowed interaction is the KPZ coupling $\sim g\tilde\theta(\nabla\theta)^2$, which can be relevant and can therefore impact the long time and distance scaling behavior. In addition, there is a notion of directionality -- the phase $\theta$ grows continuously in the direction the limit cycle is tracing out, see Fig.~\ref{fig:phase_space}. This excludes a reflection symmetry $\theta \to - \theta$, and thus the KPZ coupling is symmetry allowed and will be generated under coarse graining.
Conversely, the KPZ equation necessarily induces growth and is thus fundamentally linked to time translations in the context of broken symmetries: the KPZ nonlinearity is
forbidden for any other kind of spatial or internal
symmetry-breaking pattern, in and out of equilibrium.

\paragraph*{Scaling predictions --} This framework allows us to predict the scaling behavior in time crystals.
In low dimensions $d=1,2$, the KPZ coupling is relevant, and we expect strong fluctuations linked to the non-diffusive behavior of the chronon, measured by the phase correlator $C_\theta(t,\xx)= \expVal{(\theta(t,\xx)-\theta(0,0))^2}$.
Let us discuss how this impacts generic correlation functions. Since the limit cycle is periodic with period $T=2\pi/\Omega$, we can Fourier expand the order parameter field as $\vvarphi(t)=\sum_n \veca_n \cos(n\Omega t+\delta_n)$, and fluctuations $\theta(t,\xx)$ along the limit cycle thus lead to $\boldsymbol{\phi}(t,\xx)=\sum_n \veca_n \cos((n\Omega t+n\theta(t,\xx)+\delta_n)$. By means of a cumulant expansion, the correlation function of these fluctuations thus reads
\begin{align}
\label{eq:Cfield}
    \begin{split}
    C(t,\xx)=&\langle \boldsymbol{\phi}(t,\xx)\cdot\boldsymbol{\phi}(0,0)\rangle\\
    =&\sum_n \boldsymbol{a}_n^2 \cos(n\Omega t+\delta_n)e^{-\frac{n^2}{2} 
    \expVal{(\theta(t,\xx)-\theta(0,0))^2}_c}\\
   \sim &  e^{-\frac{1}{2}\langle(\theta(t,\xx)-\theta(0,0))^2\rangle_c},
\end{split}
\end{align}
where in the last line, we specify the leading scaling of the envelope of the correlation function. If the phase fluctuations are within the KPZ universality class, this means that

\begin{align}
\ln C(t,\xx)= -A |\xx|^{2\chi} \mathcal{C}(t/\xx^z) \Rightarrow  \left\{\begin{array}{c}
      \ln C(t,0)\sim -A t^{2\beta}   \\
     \ln  C(
       0,\xx)\sim -B |\xx|^{2\chi} \end{array} \right. ,
\label{eq:autocorr} 
\end{align}
where $\chi$ is the roughness exponent, $\beta=z\chi$ with $z$ the dynamical critical exponent, and $\mathcal{C}$ a universal scaling function~\cite{Taeuber2014,Praehofer2000,Praehofer2004,Kloss2012,Deligiannis2022}. The exponents are known exactly in one dimension $d=1$: $\beta=1/3$, $\chi=1/2$, and approximately in higher dimensions. Specifically, $\beta\approx 0.24$ and $\chi\approx 0.39$~\cite{HalpinHealy2012,Pagnani2015,Carrasco2022,Oliveira2022} in $d=2$.

Adding to these smooth long-wavelength KPZ fluctuations, there is a second nonlinear effect: The phase compactness allows for the existence of topological defects, space-time vortices and solitons in $d=1$~\cite{He2017,Vercesi2023}, and (spatial or space-time) vortices in $d=2$~\cite{Wachtel2016,Sieberer2016a,Deligiannis2022}. Once present, these defects will cause exponential decay of the correlation functions beyond a scale $L_\text{v}$ set by their mean separation. In turn, $L_\text{v} \sim e^{\Delta/\sigma}$, where $\Delta$ is the activation action of a defect, and $\sigma$ the noise level, for not too large nonlinearities. Out of equilibrium, these defects are always activated in $d=1,2$ \cite{He2017,Wachtel2016,Sieberer2016a}, i.e. $\Delta$ does not scale with system size. However, at low noise level and on intermediate scales $|\boldsymbol{x}|, t^{1/z}\ll L_\text{v}$, one can observe KPZ scaling. It is worth emphasizing that both effects eventually destroy any long-range order at large scales.

\begin{figure}
    \centering
    \includegraphics[width=8.54cm]{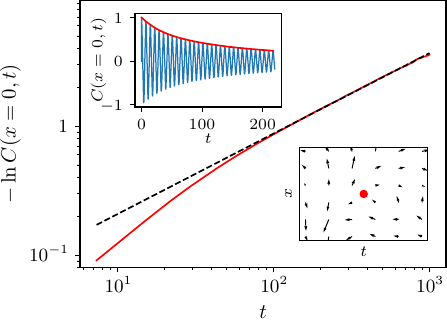}
\caption{Autocorrelation function $C(x=0,t)$ obtained by numerical simulations of Eq.~\eqref{eq:VdP} discretized for a Van der Pol chain of length $1024$, averaged over $1000$ noise realizations. The parameters are $r=-\gamma=Z_2=u=1$, $Z_1=3$, and $D=1$.
The autocorrelation function displays rapid oscillations (see inset), but its envelope (red line) shows the expected KPZ scaling  at sufficiently large scales as shown through the fit for $t \in [10^2,10^3]$ (black dotted line), which yields $\beta=0.31$, which is in very good agreement with the theoretical value. The right bottom inset shows a space-time vortex of $(\phi,\partial_t\phi)$ at higher noise, $D=2$.}
    \label{fig:1d_scaling}
\end{figure}
\begin{figure*}
    \centering
      \subfloat[\label{fig:PhaseDiagram_a}]{
    \includegraphics[width=8.54cm]{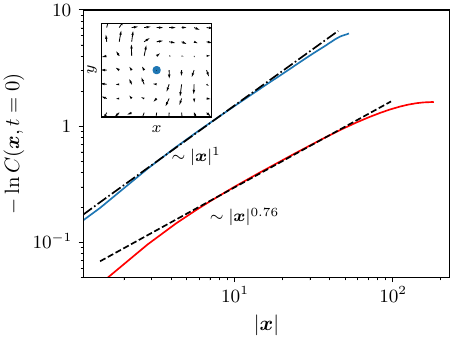}
    }
      \subfloat[\label{fig:PhaseDiagram_b}]{
    \includegraphics[width=8.54cm]{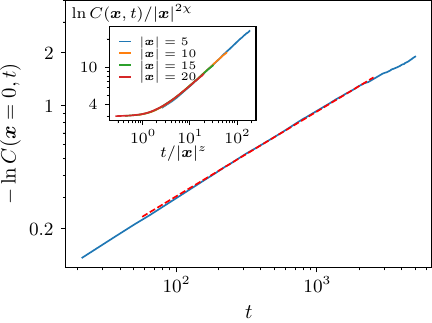}
    }
    \caption{Simulations of the Van der Pol equation in $d=2$ for a square lattice of size $256 \times 256$, and $r=-\gamma=Z_2=u=1$, $Z_1=1.25$, averaged over $1000$ realizations.  (a) The decay of the equal-time correlation displays KPZ scaling with exponent $2\chi\approx 0.76$ for $D=2$, as shown by the fit (blacked dashed line). The scaling at large $|\xx|$ is cut off by the finite system size. When $D=2.7$, the scaling is exponential, and is associated with spatial vortices, see inset for a snapshot from the corresponding simulation. (b) The autocorrelation, for $D=2$, also shows KPZ scaling and the fit (dashed line) yields $\beta=0.24$. The inset shows the correlation function for different $\xx$ plotted in a scaling form, using the fitted exponents. The scaling is consistent with Eq.~\eqref{eq:autocorr},and other KPZ simulations~\cite{Deligiannis2022,helluin2024} and theoretical prediction~\cite{Kloss2012}.  }
    \label{fig:2d_scaling}
\end{figure*}

\paragraph*{Chronons in the many-body Van der Pol system   --} As a concrete and paradigmatic example, we consider an array of classical or quantum Van der Pol oscillators in $d$ spatial dimensions~\cite{vanderPol1920,Walter2014,Walter2014a,Lee2013,Dutta2019,Arosh2021,Cabot2024}. This model has no continuous internal symmetry which could be broken spontaneously, and naively, one then might not expect any Goldstone mode to occur.  It can be considered as an effective model for generic time-translation symmetry breaking phases arising from different contexts such as nonreciprocal field theories as well as coherently driven and dissipative Bose condensates~\cite{Zelle2024}, and classical oscillators~\cite{Gutierrez2024a}. 
In the continuum limit, it is given by 
\begin{align}
\label{eq:VdP}
    \left(\partial_t^2+(2\gamma+u\phi^2-Z_1\nabla^2)\partial_t+\omega_0^2-Z_2\nabla^2\right)\phi+\xi=0,
\end{align}
where $\phi(t,\xx)$ is a real scalar field, and $\xi(t,\xx)$ a Gaussian white noise, $\langle \xi(t,\xx)\xi(t',\xx')\rangle = D \delta(t-t')\delta(\xx-\xx')$. It features limit-cycle oscillations in the regime where the damping is replaced by an antidamping $\gamma<0$.  Performing the adiabatic elimination of the transverse fluctuations and temporal averaging as anticipated above indeed yields a KPZ equation for the phase field (see SM for details~\cite{Note1})
\begin{align}
\label{eq:kpz}
    \partial_t\theta-Z\nabla^2\theta + \frac{g}{2}(\nabla\theta)^2+\xi=0,
\end{align}
where all parameters can be expressed in terms of the limit cycle  solutions of the Van der Pol oscillator. 

This explicitly bridges the microscopic physics of the spatially extended Van der Pol system to the macroscopic physics of the chronon, and thus universal scaling behavior. To confirm the predicted KPZ scaling in this system, we perform numerical simulations of the Langevin dynamics~\eqref{eq:VdP} in $d=1$ and $d=2$. We solve the stochastic differential equation discretized in real space using the Euler–Maruyama method, with a time step equal to $10^{-2}\bar{T}$. The correlation function displays the predicted KPZ scaling in time in $d=1$, Fig.~\ref{fig:1d_scaling}, which has been also observed independently in~\cite{Gutierrez2024a} as well as topological defects. In two dimensions, at low noise level we observe KPZ scaling to excellent accuracy, see Fig.~\ref{fig:2d_scaling}. At higher noise level, the topological defects proliferate, and decays are exponential. 

\paragraph*{The chronon in various platforms --} We finally point out several instances of systems where the chronon is expected to emerge, or has already been observed. 

Driven-dissipative condensates, but also classical systems close to a Hopf bifurcation, are typically described by a complex, order parameter $\psi_0=\sqrt{\rho_0}e^{-i\Omega t}$~\cite{Wouters2007,Carusotto2013,Sieberer2016,Aranson2002,Risler2005},  breaking an internal $U(1)$ symmetry next to time translations. However, both symmetries act identically on the order parameter, namely as a phase shift. The transformation generated by one of the symmetries can be compensated by a properly chosen transformation of the other. Consequently, there is only one independent generator broken, and only one Goldstone mode emerging. This mode has been theoretically demonstrated and experimentally observed to display the behavior of a compact KPZ mode \cite{Altman2015,He2015,Deligiannis2022,Fontaine2022}. But, time-translation symmetry breaking also explains and predicts the emergence of KPZ scaling in systems where such an internal continuous symmetry is broken down explicitly either fully, or to discrete ones, as in coherently driven condensates~\cite{Diessel2022} ($U(1)\searrow \mathbb{Z}_2$), or in non-reciprocal classical magnetic models~\cite{Avni2023,blom2024} ($SO(2)\searrow\mathbb{Z}_4$).

Similarly, stable traveling wave states $\expVal{\phi(t,\xx)}= \phi_0 \cos{(\Omega t -\vecq_0\cdot\xx)})$ are expected to have a single Goldstone mode, the chronon, since time and space translations again act identically on the order parameter. We thus also predict KPZ scaling, or anisotropic variants of it~\cite{Wolf1991,Note4}\footnotetext{In that case, the effective dynamics may be associated with spatially anisotropic variants of KPZ due to the preferred direction set by the spatial pattern.} generically in these systems~\cite{Cross1993,Chen2013}. Examples are encountered in magnon condensates~\cite{Rezende2009,NowikBoltyk2012}, or in  nonreciprocal systems with a conserved order parameter~\cite{Saha2020,You2020,FrohoffHuelsmann2021,Brauns2024,Suchanek2023a,Cates2024}, which were indeed predicted to display KPZ physics~\cite{Pisegna2024} in some regime, even if the presence of the additional conservation laws can play a role in general.

\paragraph*{Conclusion and outlook --} The breaking of the external continuous symmetry of time translations gives rise to a soft mode, the chronon, in turn unleashing universal scaling behavior in extended parameter regimes. Since time-translation symmetry is not affected by interactions or microscopic anisotropies potentially spoiling internal symmetries, the laid out mechanism is very robust. We have exemplified this general mechanism for the paradigmatic example of extended systems of Van der Pol oscillators both analytically and numerically. This dramatically enlarges the range of platforms offering prospects for an experimental realization of the KPZ phase. This allows for experimental tests of KPZ in higher dimensions, including the roughening transition in three dimensions.

We have focused on time-crystalline phases, where time-translation symmetry is broken down to a periodic motion, leading to an $SO(2)$ Goldstone mode. The original KPZ equation describing a growing noisy interface with collective variable $\expVal{\phi(t,\xx)} =  vt$, can likewise be embedded in the time-translation symmetry breaking scenario. 
Formally, the only difference lies in a linear or a periodic growth of a collective variable, taking values in $\mathbb{R}$ or in $\mathbb{R}/\mathbb{Z}\simeq SO(2)$. This results in fluctuations described by a compact or non-compact Goldstone mode, respectively. 

In systems realizing time-crystalline order, the respective phases may break additional internal independent symmetries, such as spin rotation symmetries in `active'~\cite{hardt2024, delSer2021,delSer2023}, non-reciprocal spin systems~\cite{Fruchart2021,Hanai2024a}, and multi-component driven condensates~\cite{Carusotto2013,Daviet2024,Weinberger2024}, or space translation symmetry~\cite{Cross1993,NowikBoltyk2012,nigro2024}.
In general, this leads to additional Goldstone modes that couple to the chronon~\cite{Zelle2024, Daviet2024}. Further soft modes can also emerge from conserved quantities~\cite{Hohenberg1977} giving rise to out-of-equilibrium hydrodynamics, particularly relevant in active matter settings. Our theory immediately predicts that the Goldstone mode of time translations will be subject to KPZ-like nonlinearities. The couplings to additional soft modes and potentially modified noise structures may lead to new universal behavior establishing novel nonthermal phases of matter with distinct scaling laws~\cite{Ertas1992,Ertas1993,Chen2013,Haldar2023,Haldar2023a}. The systematic field-theoretic study within nonequilibrium nonlinear $\sigma$-models represents an intriguing avenue for future research.

\begin{acknowledgments}
\paragraph*{Acknowledgments--}
 We thank Y. Avni, L. Canet, H. Chaté, L. Delacretaz, M. Fruchart, R. Hanai, J. Lang, A. Maitra, A. Rosch, and V. Vitelli for useful discussions. We acknowledge support by the Deutsche Forschungsgemeinschaft (DFG, German Research Foundation) CRC 1238 project C04 number 277146847. 

 \paragraph*{Data and code availability --} The code and data used to generate the figures are available open-source on Zenodo~\cite{Daviet2024b}.
 \end{acknowledgments}
 
 \FloatBarrier

\clearpage

\onecolumngrid
\setcounter{equation}{0}

\setcounter{figure}{0}
\setcounter{table}{0}
\setcounter{page}{1}
\makeatletter
\renewcommand{\theequation}{S\arabic{equation}}
\renewcommand{\thefigure}{S\arabic{figure}}
\renewcommand{\thetable}{S\arabic{table}}
\acresetall

\begin{center}
\textbf{\Large Supplemental Material for `Kardar-Parisi-Zhang scaling in time-crystalline matter'}

\end{center}

\section{Brief overview: Lindblad-Keldysh theory vs. classical stochastic systems}

Here we give a brief contextualization of Eq.~\eqref{eq:keldysh} in the main text, starting from driven open many-body quantum systems and  then connecting to classical stochastic systems~\cite{sieberer2023universality}. The former are often well described by a Lindbladian,
\begin{align}
    \partial_t \hat \rho =\int_{\xx} \big( - i [\hat h ,\hat \rho] + \sum_\alpha (\hat l_\alpha\hat \rho \hat l_\alpha^\dag - \tfrac{1}{2} \{\hat l_\alpha^\dag \hat l_\alpha,\hat \rho \} \big) ,
\end{align}
with Hamiltonian density $\hat h = \hat h [\hat \psi^\dag(\boldsymbol{x}),\hat \psi(\boldsymbol{x})]$ and Lindblad operators $\hat l_\alpha = \hat l_\alpha [\hat \psi^\dag(\boldsymbol{x}),\hat \psi(\boldsymbol{x})]$ depending on the canonical field operators $\hat \psi^\dag,\hat \psi$. For example, $\hat h$ can describe kinetic energy and local interactions of bosons, and $\hat l_1 = \sqrt{\gamma_l}\hat \psi (\boldsymbol{x}), \hat l_2 = \sqrt{\gamma_p}\hat \psi^\dag (\boldsymbol{x}),\hat l_3 = \sqrt{\kappa}\hat \psi^2 (\boldsymbol{x})$ describe single-body loss, single-particle incoherent pump, and two-body loss, respectively, with the associated rates absorbed into the operators. Having order parameter theories in mind, we restrict to bosons here.

By the usual Trotterization procedure, this theory can be reformulated as an equivalent Lindblad-Keldysh functional integral, 
\begin{eqnarray}
    Z &=& \int \mathcal D (\psi_\pm^*,\psi_\pm)\exp (i S[\psi_\pm^*,\psi_\pm]),\\\nonumber
    S&=& \int_{t,\xx} \big(\sum_\sigma \sigma (\psi_\sigma^*i\partial_t \psi_\sigma - h_\sigma ) +  \sum_\alpha (l_{+,\alpha} \ l_{-,\alpha}^* - \tfrac{1}{2} ( l_{+,\alpha}^*l_{+,\alpha} + l_{-,\alpha}^*l_{-,\alpha})\big),
\end{eqnarray}
where $h_\pm = h_\pm[\psi^*_\pm,\psi_\pm]$ etc., with the $\pm$ index tracking the left ($+$) and right ($-$) action of operators in the Lindblad equation. 

A change of basis $\psi = (\psi_+ + \psi_-)/\sqrt{2}, \tilde\psi = (\psi_+ - \psi_-)/\sqrt{2}$ introduces order parameter (or `classical') and response (or `quantum') fields. This transformation is particularly useful in generic driven open problems with (i) a finite Markovian noise level for all degrees of freedom, and (ii) gapless modes, as in the case of the Goldstone modes discussed in the main text: In this case, in the low frequency regime well below the noise level, a semiclassical limit can be taken, which amounts to a truncation up to second order in $\tilde \psi$, while the full dependence of $\psi$ is kept. In this limit, the Lindblad-Keldysh theory reduces to the Martin-Siggia-Rose-Janssen-De Dominicis (MSRJD) theory, which in turn is equivalent to stochastic Langevin equations.

The above action is expressed in terms of complex field variables. While this is natural in the context of quantum systems, clearly the action can be re-expressed in real fields according to $\psi = (q + i p)\sqrt{2}, \tilde\psi = (\tilde q + i \tilde p)\sqrt{2}$. Classical stochastic systems are often described in real coordinates from the start; for a unified approach, we work with general real vector fields in the main text, for which the Schwinger-Keldysh or MSRJD functional takes the form
\begin{align}\label{eq:keldyshapp}
    Z[\vv{\tilde{J}},\vv{J}]= \int \mathcal{D}\vphi \mathcal{D}\vv{\tilde\phi} \exp(iS[\vphi,\vv{\tilde\phi}]+i\vv{\tilde{J}} \vphi+i \vv{J} \vv{\tilde\phi}),
\end{align}
in presence of external sources.

\section{Nambu-Goldstone mode of time translations}\label{app:goldstone}

Because of time-translation symmetry, the functional integral is invariant under infinitesimal time translations $t\to t+\epsilon,\, \boldsymbol{\phi}' \to \boldsymbol{\phi}+\epsilon\partial_t \boldsymbol{\phi}$ whose only generator is $\partial_t$. From the Goldstone theorem~\cite{Schwartz2014,Dupuis2022}, we therefore get a Goldstone mode from its breaking whenever the order parameter is time-dependent~\cite{Zelle2024,Daviet2024}
\begin{align}
\label{eq:Gold_T_tt}
    \int_{t'} \left(G^R\right)^{-1}_{i,j}(t,t',\boldsymbol{q}=0)\partial_{t'}\varphi_j(t') =0,
\end{align}  
where $G^R$ is the response to the external source $J$. As long as $\partial_t\vvarphi\neq 0$, this implies that the response function possesses a pole at vanishing momentum. 
If we can go into a frame that is comoving with the limit cycle where $\vvarphi(t)=v\hat{e}_\parallel$, we have
\begin{align}
    (G^R)^{-1}_{\parallel}(\omega=0,\vecq=0)=0.
\end{align}
the Goldstone mode corresponds to a pole in the  response function at vanishing frequency and momenta. At the linear level, the existence of such a frame is ensured by the Floquet theorem.

\section{Effective Goldstone dynamics for limit cycles}\label{app:rwa}

\subsection{Symmetry based construction}

In the symmetry broken phase, time-translation symmetry is still realized, albeit in a nonlinear way. It acts in the following way on the fields: 
\begin{align}
\theta(t,\xx) \to \theta(t+t_0,\xx)+ \frac{t_0}{\bar T}, \quad \tilde \theta(t,\xx) \to \tilde \theta(t+t_0,\xx),
\end{align}
where $t_0 \in \mathbb{R}$.
Since a constant $\theta$ corresponds to a symmetry operation, the associated variation of the action has to vanish, $S[\tilde\theta(t+\theta,\xx),\varphi(t+\bar{T}\theta)]-S[\tilde\theta(t,\xx),\varphi(t)]=0$. Therefore, all terms that appear in the effective action $S_\theta$ that are linear in $\tilde \theta(t,\xx)$ have to involve at least one derivative acting on $\theta(t,\xx)$. This is analogous to  the usual shift symmetry of an internal $U(1)$ Goldstone mode. However, due to the external nature of time-translation symmetry, there is additional $\theta$ dependencies. These arise from acting with differential operators on the fluctuation around the limit-cycle stable state $\varphi(t+ \theta(t,\xx))\bar T$ and therefore always appear as $f(t+\theta\bar{T})$ for some function $f$. Clearly, this also leads to explicit time dependencies. From a symmetry perspective, this is a consequence of the fact that the combination $t+\theta(t,\xx)\bar{T}$ is also invariant under a time translation~\cite{hongo2019}.  Altogether, the action takes the form 
\begin{align}
\label{eq:S_app}
    S_\theta[\tilde\theta,\theta]=\int_{t,\xx}\tilde\theta\left(\partial_t-Z[t+\theta\bar T]\nabla^2\right)+\frac{1}{2}g[t+\theta\bar T]\tilde\theta(\nabla\theta)^2-2D[t+\theta\bar T]\tilde\theta^2,
\end{align}
But, the functions $Z,D$ and $g$ that carry these explicit time dependencies are periodic functions with the limit-cycle period $\bar T$. Considering timescales much larger $\tau\gg\bar T$, one can safely eliminate these time dependencies by averaging over one period $\bar T$. This makes the action only implicitly time-dependent, and the KPZ action of the main text is recovered. 

Some constraints apply on the response field part of the action. First, the action must vanish for $\tilde \theta=0$ to respect conservation of probability~\cite{Kamenev2011}. In addition, the action~\eqref{eq:S_app} describes the dynamics if $\theta$ is not a conserved field. Indeed, around a limit cycle, we do not expect the Goldstone fluctuations to be conserved in general, in particular in driven open systems. However, there may be special cases, where $\theta$ is subject to a conserved noise, which could lead to different variant of the KPZ universality class~\cite{Sun1989,Wolf1990,Taeuber2014}. Finally, we note that all the other operators that are allowed by symmetry are irrelevant in the low-frequency regime near the Gaussian theory, $g=0$, and they will not change the universal physics.

Because the KPZ action necessarily implies a growth, $\langle \theta \rangle \propto t$ (i.e, it generates a correction to $\bar T$), it is fundamentally associated to time-translation breaking for Goldstone modes: The KPZ nonlinearity is forbidden for any kind of spatial or internal symmetry-breaking pattern.

Finally, we emphasize that another mechanism to obtain KPZ scaling is from the equivalence to the Burgers equation~\cite{Taeuber2014}. This equation can describe, for example, the dynamics of a conserved quantity in one dimension,
\begin{align}
    \partial_t\rho -\partial_x^2\rho + \frac{\lambda}{2}\partial_x(\rho^2) +\partial_x \xi =0.
\end{align}
The KPZ equation Eq.~(8) of the main text is recovered by setting $\rho=\partial_x \theta$. It is, however, clear in that case that the growth of $\theta$ is not physical. This case is not included in our analysis, which focuses on standard Goldstone modes arising from (weak) symmetry breaking.

\subsection{Explicit construction for the Van der Pol oscillator}

We now explicitly construct the effective dynamics in a limit-cycle phase for a two-dimensional order parameter $\vphi=(\phi_1,\phi_2)$. 
The construction can equivalently be done at the level of the action, or on the equivalent stochastic equation of motion. For concreteness, we will use the Van der Pol oscillator as an explicit example
\begin{align}
\label{eq:VdPapp}
    \left(\partial_t^2+(2\gamma+u\phi^2-Z_1\nabla^2)\partial_t+\omega_0^2+\lambda\phi^2-Z_2\nabla^2\right)\phi+\xi=0,
\end{align}
where $(\phi_1,\phi_2) =(\phi,\pi=\partial_t \phi) $,but the following formulas are easily leveraged to an arbitrary case.
For negative damping $\gamma<0$, its mean-field solution is a limit cycle, which we denote $f(t)$, and time-translation symmetry is spontaneously broken. 

As shown in Eq. (3) of the main text, the idea is to work in the moving frame that locally has one of its basis vectors along the limit-cycle solution, the so-called Frenet frame. For the Van der Pol case, the tangent vector $\hat{\mathbf{e}}_{\theta}(t)$ and normal vector $\hat{\mathbf{e}}_{N}(t)$ to the limit-cycle mean-field solution $\vvarphi(t)=(f(t), \dot{f}(t))$ are given by 
\begin{align}
    \hat{\mathbf{e}}_{\theta}(t)= \frac{ \dot{\vvarphi}t)}{||\dot{\vvarphi}(t)||}=\frac{(\dot{f}(t),\ddot{f}(t))^T}{f(t)^2+ \dot{f}(t)^2}, \quad \hat{\mathbf{e}}_{N}(t)=\frac{(-\ddot{f}(t),\dot{f}(t))^T}{f(t)^2+ \dot{f}(t)^2}.
\end{align}
 The fluctuations around the limit cycle can then be parameterized as
\begin{align}
     \left(\begin{array}{cc}
         \phi(t,\xx)  \\
         \pi(t,\xx)
    \end{array}\right)= 
        \left(\begin{array}{cc}
        f(t+\theta(t,\xx)\bar T)  \\
        \dot{f}(t+\theta(t,\xx)\bar T) 
    \end{array}\right)+N(t,\xx)\hat{\mathbf{e}}_N(t+\theta(t,\xx)\bar T).
\end{align}
This ensures the gaplessness of $\theta(t,\xx)$, because a constant $\theta$ simply transports along the limit cycle (along $\hat{\mathbf{e}}_\theta$ for an infinitesimal $\theta$). For an order parameter that has more than two components, the construction can be generalized by identifying the $M-1$ dimensional hyperspace, which is orthogonal to the limit cycle at every time $t$.

This parametrization can then be inserted into the action or Langevin dynamics. Once projected onto the two vectors $\hat{\boldsymbol{e}}_\theta$ and $\hat{\boldsymbol{e}}_N$, the latter takes the form 
\begin{align}
    &\partial_t N - Z_N (t+\theta\bar{T}) \nabla^2 N + m_N(t+\theta\bar{T}) N +  Y_N(t+\theta\bar{T}) \nabla^2 \theta+g_N(t+\theta\bar{T}) (\nabla\theta)^2 +D_N(t+\theta \bar{T})\xi_N= 0,\\
    &\partial_t \theta - Z_\theta(t+\theta\bar{T}) \nabla^2 \theta +  Y_\theta(t+\theta\bar{T}) \nabla^2 N + m_\theta(t+\theta\bar{T}) N+  g_\theta(t+\theta\bar{T}) (\nabla\theta)^2+ D_\theta(t+\theta \bar{T})\xi_\theta=0,
\end{align}
where we neglected all nonlinear terms in $N$, noticing that its fluctuations are gapped, while the gapless Goldstone nature of $\theta$ is evident. 
Again, all the explicit $t$ dependencies are functions of $f(t)$ and its derivatives. In particular, we have 
 \begin{align}
 \label{eq:coeff_vdp}
     Z_\theta=\bar{T} \ddot{f}(t)\frac{(Z_2\dot{f}(t)+Z_1\ddot{f}(t))}{\dot{f}(t) ^2+{\ddot{f}(t)}^2}, \quad \text{and } g_{\theta}=-\bar{T}^2 \frac{\ddot{f}(t)(Z_2\ddot{f}(t)+Z_1\dddot{f}(t)}{{\dot{f}(t)} ^2+{\ddot{f}(t)}^2}.
 \end{align}
 
Now, the Goldstone mode fluctuates on all timescales and we are in particular interested in long times $t \gg \bar T$, where $f(t)$ oscillates very quickly. It is therefore justified to average the right hand side over time following the average method~\cite{Verhulst1996}. Such a procedure is described in the noiseless case in, e.g.,~\cite{Cross1993}, and leads to a so-called phase equation. We point out that the KPZ nonlinearity was already been shown to be associated to temporal instabilities in the noiseless case~\cite{Cross1993}. Neglecting the gapped fluctuations of the phase, this procedure yields the KPZ equation~\eqref{eq:kpz} of the main text with coefficients $g=\bar{T}^{-1}\int_0^T dt g_\theta(t) $ and $Z=\bar{T}^{-1}\int_0^T dt Z_\theta(t)$. To compute the corrections coming from the gapped degree of freedom, we can use the first equation to adiabatically eliminate $N$, i.e., solving the first equation for $N(\theta)$. This is equivalent to performing the Gaussian integral over $N$ and its response field in the path-integral formalism.  

This derivation shows how KPZ dynamics for a gapless chronon emerges in a limit-cycle phase in any dimensions without broken internal continuous symmetries: it solely depends on the breaking of time-translation symmetry. 
\end{document}